\documentclass[aps,prd,twocolumn,nofootinbib,superscriptaddress]{revtex4-1}

\usepackage{amsmath}
\usepackage{amssymb}
\usepackage{graphicx}

\newcommand{\be}{\begin{equation}}
\newcommand{\ee}{\end{equation}}
\newcommand{\bea}{\begin{eqnarray}}
\newcommand{\eea}{\end{eqnarray}}

\newcommand{\gsim}{\lower.7ex\hbox{$\;\stackrel{\textstyle>}{\sim}\;$}}
\newcommand{\lsim}{\lower.7ex\hbox{$\;\stackrel{\textstyle<}{\sim}\;$}}
\newcommand{\ksg}{\kappa_s}
\newcommand{\ksw}{\kappa_w}
\newcommand{\bs}{b}

\newcommand{\ATLAS}{ATLAS}
\newcommand{\CMS}{CMS:2015dxe}

\begin{document}


\title{First interpretation of the 750 GeV di-photon resonance at the LHC}

\author{Stefano Di Chiara}
\affiliation{National Institute of Chemical Physics and Biophysics, R\"avala 10, 10143 Tallinn, Estonia.}
\author{Luca Marzola}
\affiliation{National Institute of Chemical Physics and Biophysics, R\"avala 10, 10143 Tallinn, Estonia.}
\affiliation{Institute of Physics, University of Tartu, Ravila 14c, 50411 Tartu, Estonia. }
\author{Martti Raidal}
\affiliation{National Institute of Chemical Physics and Biophysics, R\"avala 10, 10143 Tallinn, Estonia.}
\affiliation{Institute of Physics, University of Tartu, Ravila 14c, 50411 Tartu, Estonia. }

\date{\today}

\begin{abstract}
We scrutinise the evidences recently reported by the ATLAS and CMS collaborations for compatible 750~GeV resonances which appear in
the di-photon channels of the two experiments in both the 8~TeV and 13~TeV datasets. 
Similar resonances in di-boson, di-lepton, di-jet and $t\bar t$ final states are instead not detected.
After discussing the properties and the compatibility of the reported signals, we study the implications on the physics beyond the Standard Model with particular emphasis on possible scalar extensions of the theory such as singlet extensions and the two Higgs doublet models. We also analyse the significance of the new experimental indications within the frameworks of the minimal supersymmetric standard model and of technicolour models. Our results show that a simple effective singlet extension of the SM achieves phenomenological viability with a minimal number of free parameters. The minimal supersymmetric model and the two Higgs doublet model, on the other hand, cannot explain the 750~GeV di-photon excess. Compatibility with the observed signal requires the extension of the particle content of these models, for instance by heavy vector quarks in the case of the two Higgs doublet model.

\end{abstract}

\maketitle


\section{Introduction} 
\label{sec:Introduction}

The discovery of the Higgs boson~\cite{Aad:2012tfa,Chatrchyan:2012xdj}, possibly the first spin zero elementary particle observed in Nature, 
raised the crucial issue concerning the existence of possibly several scalar particles with masses much below any supposed cut-off scale of a given theory, such as the Planck scale. The detection of a light scalar sector would, in fact, allow us to discriminate between the theories beyond the Standard Model (SM) which protect the electroweak scale from the influence of the high-energy cut-off, such as supersymmetry or compositeness, and the scenarios supported by selection mechanisms or landscape arguments which disfavour the existence of these particles.

Recently, both the ATLAS~\cite{\ATLAS} and CMS~\cite{\CMS} experiments at the LHC have reported an excess of events in the di-photon channel 
associated to an invariant mass of about 750~GeV. Given the energy resolutions of the experiments, the signal events seem consistent with each other, implying an evidence for new physics with a global statistical significance that certainly exceeds the $3~\sigma$ level. From a theoretical point of view, because spin one particle decays to di-photon final states are forbidden by the Landau-Yang theorem, the possible candidates for the new resonance must have either spin zero or two.
However, in both the cases, the fact that no excesses have been reported for comparable energies in complementary channels as the di-jet~\cite{Khachatryan:2015dcf} and $t\bar t$~\cite{Khachatryan:2015sma,Aad:2015fna}, and neither in di-boson~\cite{Aad:2015owa} nor di-lepton~\cite{Aad:2015ufa} final states, poses a clear challenge to the possible interpretations of the di-photon excess within models of new physics. 

In this work, after discussing the consistency of the LHC di-photon resonances detected by the two experiments, we interpret the signal in terms of a new hypothetical scalar particle and investigate the mentioned experimental hints within an effective field theory that models a possible singlet extension of the SM, as well as within the four flavour conserving Two Higgs Doublet Models (2HDM). We pay particular attention also to the minimal supersymmetric standard model (MSSM), study in detail a simple 2HDM extension featuring two heavy vector-like quarks, and comment, for completeness, on the possibilities offered by composite resonances. 

Our results show that the LHC di-boson excess is indeed compatible with all the mentioned models but the 2HDM, including its supersymmetric UV completion, the MSSM, which are strongly disfavoured by the LHC upper constraints on the $pp\rightarrow H\rightarrow t\bar{t}$ cross section.

\section{Consistency of the signal } 
\label{sec:H}

Recently the ATLAS and the CMS collaborations presented their results for searches of resonances in the di-photon channel analyzing respectively 3.2~fb$^{-1}$ and 2.6~fb$^{-1}$ of data collected at a 13~TeV collision energy. Both the experiments observe an excess in the di-photon signal peaked at 747~GeV~\cite{\ATLAS} and 
750~GeV~\cite{\CMS} with local significances of 3.6 and $2.6~\sigma$ in ATLAS and CMS, respectively. 
In addition to that, the CMS collaboration presented the combined results that include  19.7~fb$^{-1}$ of published data taken at 8~TeV~\cite{Khachatryan:2015qba}, which exhibits an excess
at the same energy and consequently enlarges the local significance of the signal to the $3.1~\sigma$ level. The ATLAS collaboration did not present the corresponding combination since the relative 
Run 1 analysis extends only to 600~GeV of invariant mass. Nevertheless, during the presentation of the new results, the speaker~\cite{\ATLAS} remarked that the two ATLAS datasets are consistent with each other.
The uncertainty in the photon energy determination at 750~GeV is of order ${\cal O}(1)\%$ in both the experiments, 
depending on whether one of the photons is detected in the barrel or in the endcap. 
Therefore, within the quoted energy uncertainty, the signals detected by the two experiments are compatible with each other and can originate from the decays of a new particle.

In light of this, regarding the two datasets as statistically independent and barring systematic errors allows us to make a first, naive, combination of the two signals that rejects the SM background hypothesis at the local $4.5~\sigma$ level. 
Clearly the corresponding global significance is to be diminished by the look-elsewhere effect. However, in the combination of two independent measurements the signal of one experiment could be used to determine the signal region, compensating by net look-elsewhere effect, in a way that the signal detected by the other experiment then acquires a global significance. This implies that the total global significance of the LHC di-photon excess at 750~GeV invariant mass exceeds the $3~\sigma$ level and
should be considered as a strong evidence for new physics.
  
Breaking down the signal, we see that the most significant excess in the di-photon invariant mass spectrum observed by CMS for barrel-barrel events in the combined 8+13 TeV analysis is at around 750~GeV, suggesting a production cross section times branching ratio of

\be
\sigma^{\rm CMS}_{pp\to H}BR^{\rm CMS}_{H\to\gamma\gamma}=4.47\pm 1.86~{\rm fb},
\ee
obtained by the combination, properly scaled, of 8 and 13 TeV data.  
  
As for the ATLAS signal, the most significant excess in the diphoton invariant mass spectrum is observed around 747~GeV. The difference between the number of events predicted by SM and the data is equal to
\be
\Delta N=13.6\pm3.69,
\ee
which, given the efficiency and acceptance values for the mentioned invariant mass and the integrated luminosity, corresponds to a cross section of:
\be
\label{eq:exp_Sigmabr}
\sigma^{\rm ATLAS}_{pp\to H}BR^{\rm ATLAS}_{H\to\gamma\gamma}=\frac{\Delta N}{\epsilon\times {\cal L}}=\frac{13.6\pm3.69}{0.4\times 3.2}~{\rm fb}=10.6\pm2.9~{\rm fb}.
\ee

We remark that the quoted values of the cross section are compatible with each other at the $1.8~\sigma$ level.\footnote{We point out that only the ATLAS experiment reported a first estimate of the resonance width of about 45~GeV. The CMS collaboration was not able to resolve the width even though 20~GeV bins were employed in the analysis. Given that the uncertainty on the ATLAS width estimate is unknown, and likely large, we chose to disregard the constraints brought by this estimate in our analysis.} Were these excesses generated by a Higgs boson with mass equal to 750~GeV, its signal strength compared to a SM Higgs with the same mass in the narrow-width approximation, defined for a given scalar $\varphi$ by
\be\label{mugammaH}
\mu_{X,\varphi}=\frac{\sigma_{pp\rightarrow \varphi}{\textrm{Br}}_{\varphi\rightarrow X \bar{X}}}{\sigma_{pp\rightarrow \varphi}^{\textrm{SM}} \textrm{Br} ^{\textrm{SM}}_{\varphi\rightarrow X \bar{X}}},
\ee
 would be equal to
\be\label{mugammaHAC}
\mu_{\gamma,H}^{\rm ATLAS}=(6.4\pm1.7)\times 10^4 ,\quad \mu_{\gamma,H}^{\rm CMS}=(2.7\pm1.1)\times 10^4\ .
\ee
In the following we use this result as a basis for our computation and refer to a combined cross section 
\be\label{scomb}
\hat{\sigma}_{pp\to H}\widehat{BR}_{H\to\gamma\gamma} = 6.26 \pm 3.32~{\rm fb}~.
\ee
\section{Effective  singlet extensions} 
\label{sec:H}

\begin{figure}[h]
  \centering
    \includegraphics[width=.9\linewidth]{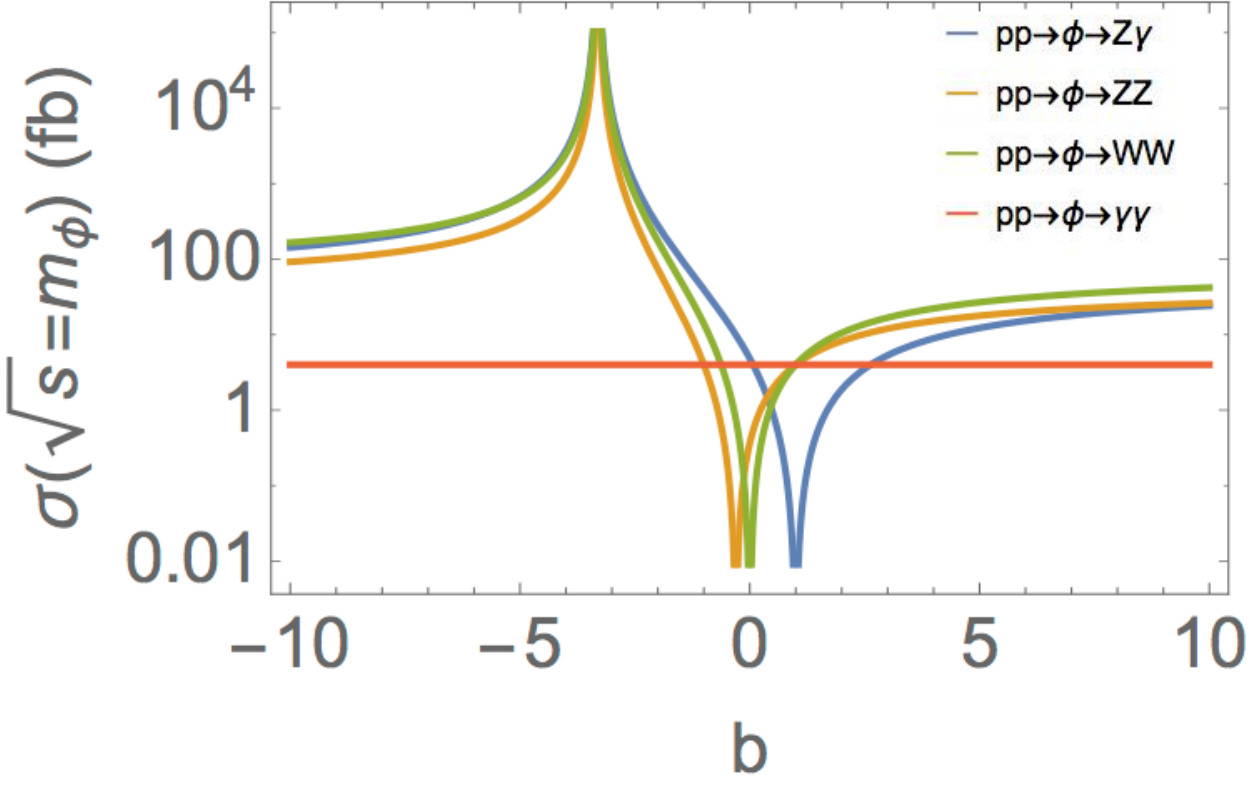}
  \caption{The production cross section times branching ratios of electroweak gauge bosons for an on-shell 750 GeV pseudoscalar or scalar mediator as a function of the parameter $b$ of Eq.~\eqref{Lsinglet}.}
  \label{fig:spin0}
\end{figure}

We start our analysis by extending the SM with a singlet spin-0 particle, $\phi$, that we assume for definiteness to have odd parity.  Analogous results will however hold for the scalar case. Barring a portal coupling $\lambda_p H^2 \phi^2$, strongly constrained by the SM Higgs couplings measured at LHC \cite{Cheung:2015dta},
and by assuming that the contact between $\phi$ and the SM gauge boson is provided only by heavy particles which transform non-trivially under the SM symmetry group, we can write the effective interaction Lagrangian \cite{Coleppa:2012eh}

\begin{align}
\label{Lsinglet}
\mathcal{L_I}
 &=
 \ksg\frac{\alpha_s}{4\pi v}\phi \sum_a G^{a}_{\mu\nu}\tilde{G}^{a\mu\nu}
 + \\ \nonumber
 &+\ksw\frac{\alpha}{4\pi v}\phi\left[B_{\mu\nu}\tilde{B}^{\mu\nu}
 	+\bs \sum_c W^{c}_{\mu\nu}\tilde{W}^{c\mu\nu} \right],
\end{align}

where $\ksg$, $\ksw$, and $\bs$ are free parameters and the tilded tensors represent the dual field strength tensors. Notice that whereas reproducing the di-photon signal bounds a combination of the former quantities, the cross section times branching ratio into a di-gluon final state depends solely on $\ksg$\footnote{ Because of the hierarchy in the coupling constants we expect that $\Gamma_{\phi\to \gamma\gamma} \ll \Gamma_{\phi\to gg}\simeq \Gamma_{tot}$ hold on most of the available parameter space.}. The value of this parameters is consequently bounded by the measurements of $\sigma(pp\to \phi \to gg)$, however, our Lagrangian in Eq.~\eqref{Lsinglet} allows us to match the observed $\sigma(pp\to\phi\to\gamma\gamma)$ irrespectively of the value assigned to $\ksg$ by simply adjusting $\ksw$ as required. The ratios between the branching ratios into the electroweak gauge bosons are instead regulated solely by $b$. In this case, by using the reference di-photon cross section value quoted in Eq.~\eqref{scomb}, we can infer the production cross section times branching ratio in the remaining electroweak bosons by simply multiplying the cross section for the relevant ratio of branching ratios. In Fig.~\ref{fig:spin0} we demonstrate the dependence of the electroweak gauge bosons production from the parameter $b$ in the approximation of massless outgoing particles.

As we can see, within this model it is clearly possible to suppress the signal in the diboson and $\gamma Z$ channels as required by the ATLAS and CMS signals by simply requiring that $|b|<1$. 

Direct couplings of the pseudoscalar $\phi$ to SM fermions $f$ can be written in the following fashion
\be
{\cal L}_{\phi f\bar{f}}=-i \kappa_f \frac{y_f}{\sqrt{2}}\phi\bar{f}\gamma^5 f
\ee
where we take the Yukawa coupling $y_f$ equal to its SM value rescaled by a factor $\kappa_f$, in agreement with the Minimal Flavor Violation (MFV) framework \cite{D'Ambrosio:2002ex}. However,  given the lack of signals for $\phi$ in the $\tau\bar{\tau}$ and dilepton channel, we argue that $\kappa_f\ll 1$ must hold for every SM fermion and consequently disregard these interactions in our analysis.

We conclude this section by remarking that a singlet scalar, coupling to SM vector bosons via an effective Lagrangian as in Eq.~\eqref{Lsinglet} where the dual field strength tensor are replaced by the ordinary strength tensors, would present the same cross section times branching ratios as those shown in Fig.~\ref{fig:spin0}.

\section{Two Higgs Doublet models} 
\label{sec:2HDM}
In the 2HDM \cite{Branco:2011iw} the physical heavy scalar $H$ can have couplings to fermions that are greatly enhanced, compared to their SM values, by the coupling coefficients:
\begin{center}\label{Hcc}
\begin{tabular}{lcccc}
  & Type I & Type II & Type III & Type IV 
  \\
  $a^H_d$  
    & $\sin \alpha/\sin \beta$ 
    & $-\cos \alpha/\cos \beta$ 
    & $\sin \alpha/\sin \beta$
    & $-\cos \alpha/\cos \beta$
  \\
  $a^H_l$  
    & $\sin \alpha/\sin \beta$ 
    & $-\cos \alpha/\cos \beta$ 
    & $-\cos \alpha/\cos \beta$
    & $\sin \alpha/\sin \beta$
\end{tabular}
\end{center}
with type independent coupling coefficients for upper EW component quarks and $W$ and $Z$ gauge bosons
\be\label{uVcoups}
a^H_u=  \sin \alpha/\sin \beta\ ,\quad a^H_V=\cos(\beta-\alpha)\ .
\ee
The physical spectrum of 2HDM features also a charged Higgs $H^\pm$ and a pseudoscalar $A$. For a heavy Higgs mass $m_H>600$~GeV the model is already in the decoupling regime \cite{Altmannshofer:2012ar}, in which $H,H^\pm$, and $A$ have similar masses
\be\label{decrM}
m_A^2 = m_H^2 + O(\lambda_i v^2) = m_{H^\pm}^2 + O(\lambda_i v^2) ~,
\ee
and the mixing angles are related by
\be\label{decrab}
\alpha=\beta-\pi/2+ O(\lambda_i v^2/m_A^2) ~,
\ee
with the quartic couplings $\lambda_i$ constrained by perturbativity to values of $O(1)$. There is therefore the concrete possibility that $A$ and $H$ are too close in mass to be resolved as separate resonances, at least at the present level of accuracy, in which case the observed excess should be ascribed to both physical states. Indeed this could explain also the large width of the signal observed at ATLAS \cite{ATLAS}. For these reasons, in the following we consider $A$ and $H$ to be degenerate in mass and add their separate contributions to the di-photon decay rate. The pseudoscalar couplings, compared to a SM Higgs, are rescaled by the following coupling coefficients
\begin{center}\label{Acc}
\begin{tabular}{lcccc}
  & Type I & Type II & Type III & Type IV 
  \\
  $a^A_u$  
    & $1/\tan \beta$ 
    & $1/\tan \beta$ 
    & $1/\tan \beta$
    & $1/\tan \beta$
  \\
  $a^A_d$  
    & $-1/\tan \beta$ 
    & $\tan \beta$ 
    & $-1/\tan \beta$
    & $\tan \beta$
  \\
  $a^A_l$  
    & $-1/\tan \beta$ 
    & $\tan \beta$ 
    & $\tan \beta$
    & $-1/\tan \beta$
\end{tabular}
\end{center}
Given the size of the $\mu_{\gamma,H}$ signal strength, we expect the signal be generated at one loop by a charged particle with a large coupling coefficient. The $H^\pm$ coupling to $H$, unlike the corresponding fermions couplings, lacks an enhancement or suppression factor; furthermore its contribution to the di-photon decay amplitude is roughly 1/4 of the fermion ones. The contribution of $H^\pm$ to the di-photon effective coupling is therefore marginal and will be neglected in the present analysis. Moreover, because of Eqs.~(\ref{uVcoups},\ref{decrab}), the $H$ couplings to $W$$W$ and $Z$$Z$ are very small in the decoupling regime, though for completeness we still include them in our computation. In the same regime the contribution of the $H\rightarrow h h$ and $A\rightarrow h Z$ channels becomes negligibly small \cite{Djouadi:2005gj}, and for this reason we do not include it in the present analysis.

We determine the values of the mixing angles $\alpha$ and $\beta$ by performing a fit to the signal strengths, defined in Eq.~\eqref{mugammaH}, by minimizing
\be
\chi^2=\sum_i \left(\frac{\mu_i^{exp}-\mu_i^{th}}{\sigma_i^{exp}}  \right)^2,
\ee
where $\mu_i^{exp}$ and $\sigma_i$ are the experimental values of $\mu_{\gamma,H}$, Eq.~\eqref{mugammaHAC}, and $\mu_{\gamma,h}$, $\mu_{Z,h}$, $\mu_{W,h}$, $\mu_{b,h}$ $\mu_{\tau,h}$, with their respective uncertainties, as measured at ATLAS and CMS \cite{ATLAS2015044,CMS:2015kwa}, while $\mu_i^{th}$ are the 2HDM predictions obtained by rescaling the production cross sections and branching ratios of a 750~GeV SM Higgs, reported in \cite{Dittmaier:2011ti,Dittmaier:2012vm,Heinemeyer:2013tqa}, each with its corresponding coupling coefficient.\footnote{All the formulas necessary to perform the fit can be found for example in \cite{Alanne:2013dra}.}

The value of the minimum $\chi^2$ per degree of freedom (d.o.f.), as well as the corresponding $p$-value and $H$ coupling coefficients for each 2HDM are
\begin{center}
\begin{tabular}{lcccc}
  & Type I & Type II & Type III & Type IV 
  \\
$\chi^2/d.o.f.$  
    & 0.95 
    & 0.78 
    & 0.95
    & 0.83
  \\
  $p$-value  
    & 49\%
    & 64\%
    & 48\%
    & 60\%
   \\
  $a_u^H$  
    & -16 
    & -16 
    & -16
    & -16
  \\
  $a_d^H$  
    & -16 
    & 0.07 
    & -16
    & 0.07
  \\
  $a_l^H$  
    & -16 
    & 0.07 
    & 0.07
    & -16
\end{tabular}\label{chi2p}
\end{center}
with the same mixing angles
\be\label{optmix}
\alpha=-1.51\ ,\quad\beta=0.063
\ee 
for every 2HDM type. For such mixing angles the $A$ coupling coefficients to fermions are numerically identical to the $H$ ones. The optimal mixing angles in Eq.~\eqref{optmix} imply a negligible coupling to vector bosons and a large enhancement of the $H$ coupling to upper EW component quarks, Eqs.~\eqref{uVcoups}, as compared to that of the 125~GeV Higgs $h$. The Type II 2HDM achieves the best fit. For comparison the SM results are
\be
\chi^2/d.o.f.=2.33\ ,\quad p=1\%~.
\ee
According only to these goodness of fit results, the 2HDMs would represent a valid explanation of the 750~GeV resonance observed at LHC, while the SM would be ruled out at the 95\% CL. However, we still have to impose the stringent constraints on the partial decay widths of the scalar resonance to SM fermions discussed below.

\begin{figure*}[htb]
\begin{center}
\includegraphics[width=0.40\textwidth]{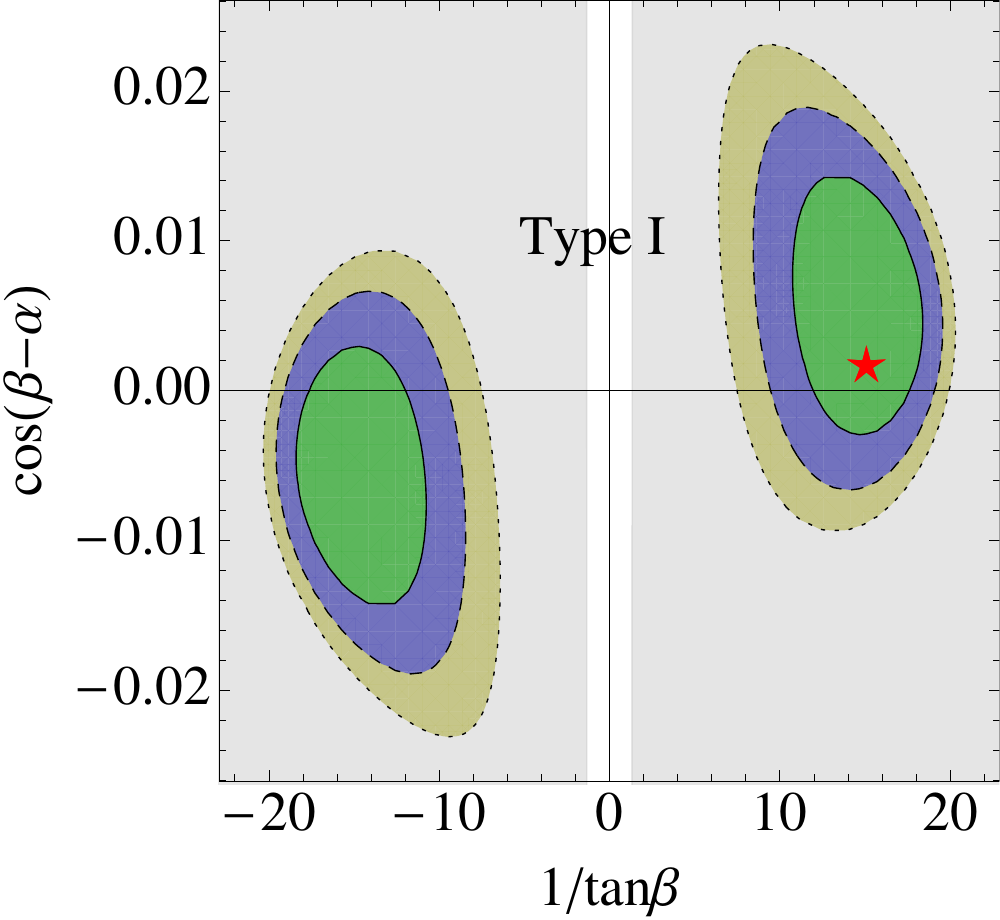}\hspace{0.5cm}
\includegraphics[width=0.40\textwidth]{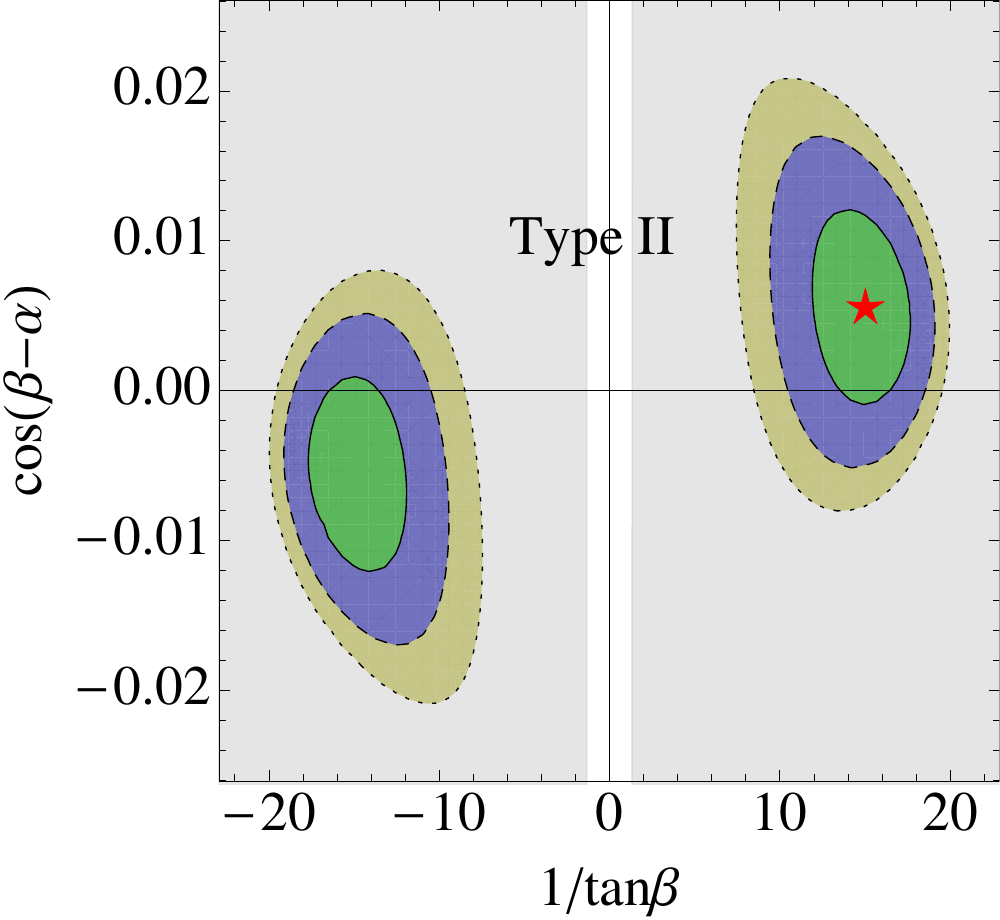}
\includegraphics[width=0.40\textwidth]{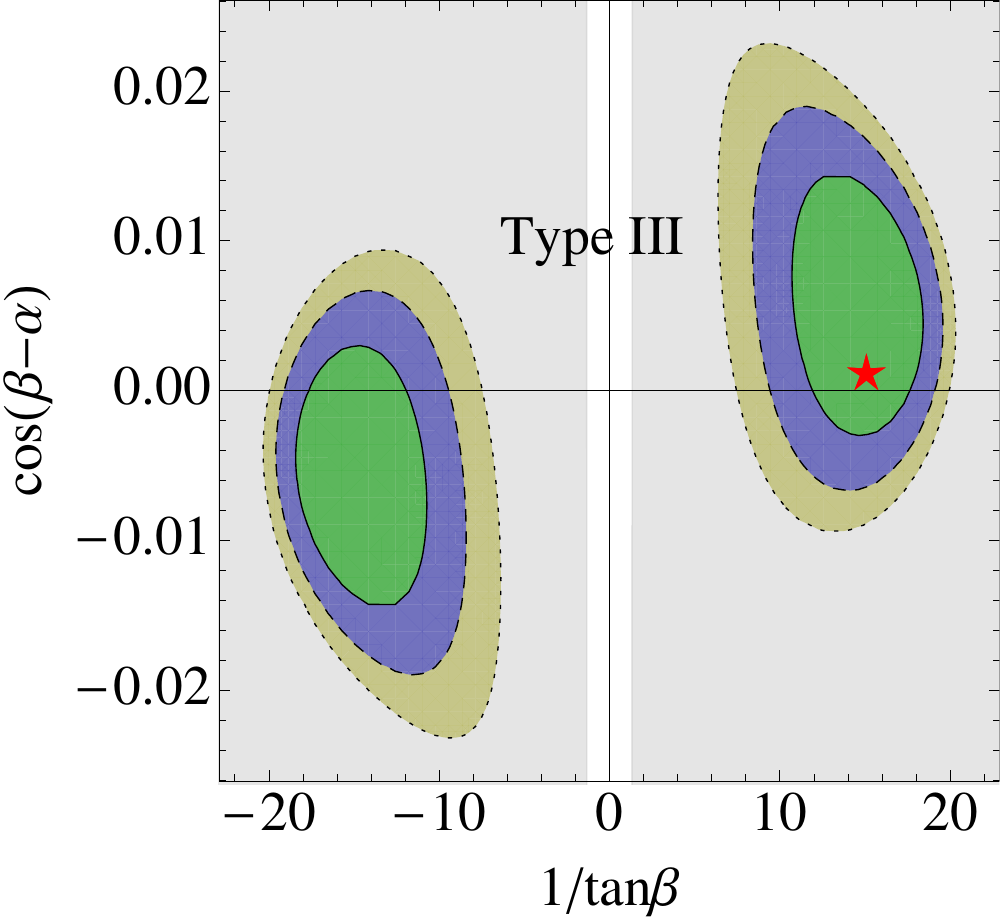}\hspace{0.5cm}
\includegraphics[width=0.40\textwidth]{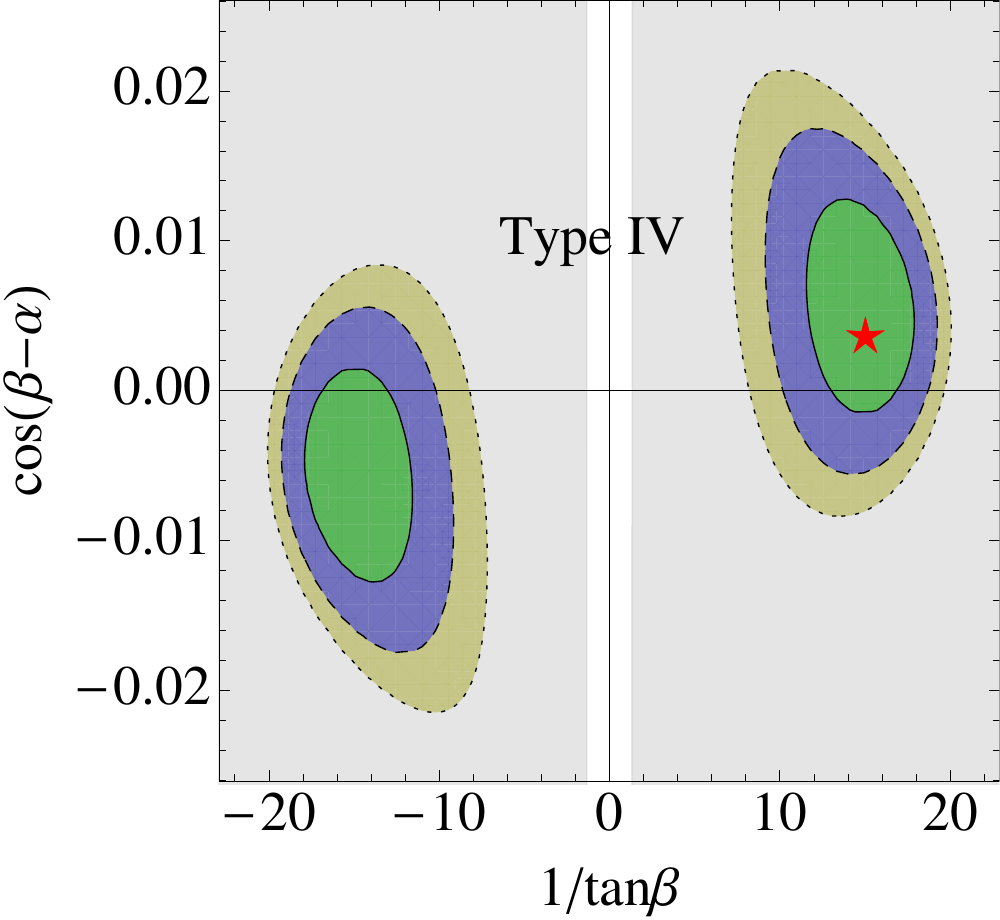}
\caption{68\%, 95\%, and 99\% CL contour plots in $1/\tan\beta$ and $\cos(\beta-\alpha)$ plane for all 2HDMs, with the shaded region excluded by the $t\bar{t}$ experimental constraint \cite{Aad:2015fna}. In each plot the red star represents the point of minimum for $\chi^2$.}
\label{fig:tbcbaTI&II&III&IV}
\end{center}
\end{figure*}

The couplings to lower component quarks and leptons are Type dependent, and for the optimal mixing angles would be suppressed in the Type II 2HDM: this model is therefore consistent with the current absence of a signal in the $WW$, $ZZ$, $\tau\bar{\tau}$, $b\bar{b}$ decay channels of $H$ at 8~TeV. On the other hand the constraint on the $t\bar{t}$ channel \cite{Aad:2015fna} needs to be imposed explicitly, given the large coupling of the $750$~GeV scalar and pseudoscalar to $t$. In the region selected by Eq.~\eqref{optmix} we can neglect, in first approximation, all the decay channels to SM particles but $t$ and gluons, and express the 8~TeV constraint \cite{Aad:2015fna} on the $pp\rightarrow H \rightarrow t\bar{t}$ cross section in terms of the SM quantities as
\bea\label{ttXS}
680~{\rm fb}&>&\sigma_{pp\rightarrow \varphi\rightarrow t \bar{t}}\sim \sigma^{\rm SM}_{gg{\rm F}}{a_t^2 \textrm{Br}}^{\rm SM}_{\varphi\rightarrow t \bar{t}}\left[ \frac{1}{{\textrm{Br}}^{\rm SM}_{\varphi\rightarrow g g}+{\textrm{Br}}^{\rm SM}_{\varphi\rightarrow t \bar{t}}}\right.\nonumber\\ 
&+&\left.\frac{\kappa_A}{\kappa_A{\textrm{Br}}^{\rm SM}_{\varphi\rightarrow g g}+{\textrm{Br}}^{\rm SM}_{\varphi\rightarrow t \bar{t}}} \right]\,,
\eea
where $gg{\rm F}$ stands for ``gluon-gluon fusion'', and $\kappa_A\sim1.41$ is the pseudoscalar decay rate to two gluons normalised to the $H$ one and both calculated for unitary $a_t$, with
\be
a_t  \equiv a_u^H\sim a_u^A=1/\tan\beta\, .
\ee
By using the values provided in \cite{Heinemeyer:2013tqa} for the SM quantities appearing in Eq.\eqref{ttXS} we obtain the constraint 
\be\label{atc}
| a_t|<1.34\,.
\ee 
In Fig.~\ref{fig:tbcbaTI&II&III&IV} we show the 68\%, 95\%, and 99\% CL contour plots of $1/\tan\beta\sim a_t$ vs $\cos(\beta-\alpha)=a^H_V$ for all the 2HDMs, with the shaded region excluded by Eq.~\eqref{atc}.
Evidently, all the 2HDM are in strong tension with the $t\bar{t}$ experimental constraint \cite{Aad:2015fna}. Nevertheless, we point out that such a bound can be easily circumvented  by adding new charged and colored particles which mediate the loop interactions of $H$ and $A$ necessary to reproduce the diphoton excess. In the next section we examine a specific case where these new particles are scalars, the stops in MSSM, while in Section~\ref{sec:2HDMvf} we study the 2HDM extended by new, vector-like quarks. To conclude, we remark that the perturbativity of the model also results in a bound that disfavour the 2HDM due to the implied $t\bar t$ coupling. We find, however, that this bound is less severe than the one implied by the observation of the $t \bar t$ channel at the LHC and, consequently, opted to neglect it.  

\section{MSSM} 
\label{sec:MSSM}
\begin{figure*}[htb]
\begin{center}
\includegraphics[width=0.40\textwidth]{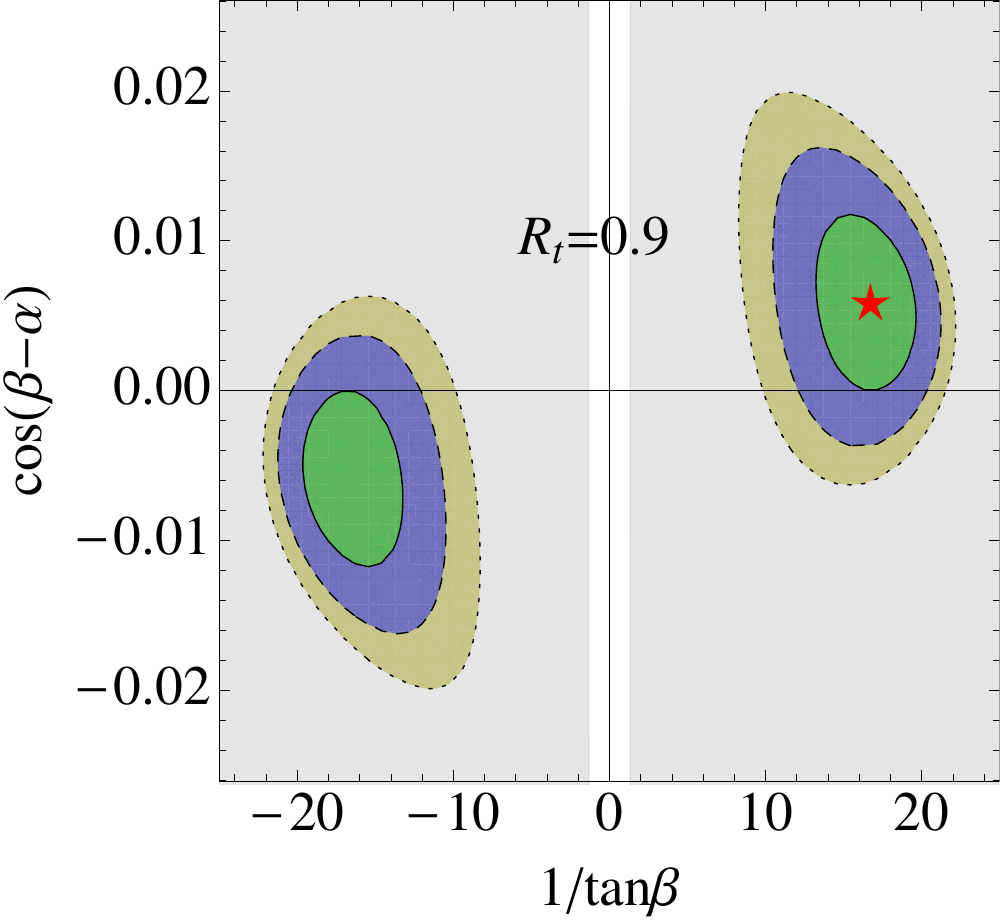}\hspace{0.5cm}
\includegraphics[width=0.40\textwidth]{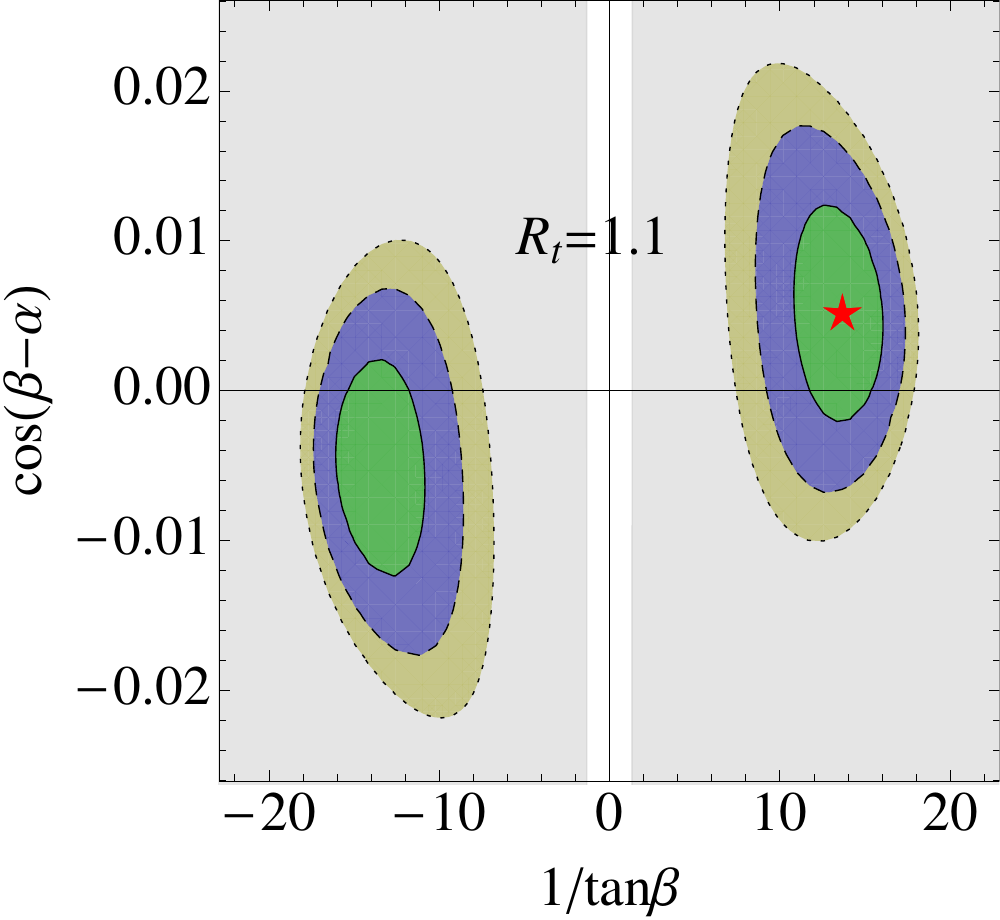}
\caption{68\%, 95\%, and 99\% CL contour plots in $1/\tan\beta$ and $\cos(\beta-\alpha)$ plane for MSSM with $R_t=0.9 (1.1)$, left (right) panel, with the shaded region excluded by the $t\bar{t}$ experimental constraint \cite{Aad:2015fna}. In each plot the red star represents the point of minimum for $\chi^2$.}
\label{3}
\end{center}
\end{figure*}

The low energy limit of MSSM corresponds to the Type II 2HDM. The most relevant correction to a Higgs decay to di-photon comes from the stop contribution, which can be expressed as a rescaling of the top coupling coefficients to both the light and heavy Higgses, respectively $h$ and $H$:
\be\label{Rt}
a_t^{\prime h/H}=R_t a_t^{h/H},\ R_{t} = 1+\frac{m_t^2}{4}\left[\frac{1}{m_{\tilde{t}_1}^2}+
\frac{1}{m_{\tilde{t}_2}^2}-\frac{X_t^2}{m^2_{\tilde{t}_1} m^2_{\tilde{t}_2}}\right],
\ee
with the stop mixing parameter
\be
X_t=A_t-\mu/\tan\beta .
\ee
Because of the tree level constraint on the light Higgs mass
\be
m_h<m_Z |\cos(2\beta)|,
\ee
the value of $\tan\beta$ is constrained in MSSM to be roughly larger than 5, for which value the stop mixing should be close to maximal
\be
\frac{X_t^2}{m^2_{\tilde{t}_1} m^2_{\tilde{t}_2}}\sim 6~.
\ee
In Fig.~\ref{3} we show the 68\%, 95\%, and 99\% CL contours in the $1/\tan\beta$ and $\cos(\beta-\alpha)$ plane for MSSM with $R_t=0.9~(1.1)$, left (right) panel. The shaded area is excluded by the constraint in Eq.~\eqref{atc}. In each plot the red star represents the point of minimum for $\chi^2$, which is characterised by a value of $\tan\beta$ too small to generate a Higgs mass of 125~GeV. While the local minima close to $\tan\beta=4$ satisfy the $t\bar{t}$ experimental constraint \cite{Aad:2015fna}, they produce a $p$-value equal to 1\% and, therefore, are strongly disfavoured. This is because a large $\tan\beta$ suppresses the top coupling to $H$ and $A$, while enhances the coupling to bottom and $\tau$, which have too small Yukawa couplings to produce the signal strength enhancement required by $\mu_{\gamma,H}$, Eq.~\eqref{mugammaHAC}. Possible corrections from the $H$ coupling to the charginos, generated by the wino, should be small given the small coupling of $H$ to $W$, proportional to $\cos(\beta- \alpha)$, as shown in Fig.~\ref{3}.

\section{2HDM extended by vector-like quarks} 
\label{sec:2HDMvf}
In this section we consider the type I 2HDM extended by two vector-like quarks $Q$ and $U'$. The charges and $\mathbb{Z}_2$ parity of the new particles, as well as those of the scalars, are given in Table~\ref{4thgen}, while the $\mathbb{Z}_2$ parity is taken positive (negative) for left(right)-handed SM fermions. The choice of the $\mathbb{Z}_2$ parity assignments ensures MFV \cite{D'Ambrosio:2002ex}. We remark that, on general grounds, models presenting  extra scalars which couple to heavy vector-fermions find a justification as low energy limits of string-inspired supersymmetric models \cite{Cvetic:2015vit,Dev:2015vjd,King:2016wep,Karozas:2016hcp}.
\begin{table}[htp]
\caption{Scalar and vector-like fermion content of the model.}
\begin{center}	
\begin{tabular}{|ccccc|}
\hline
Field & $SU(3)_c $ & $SU(2)_L$ & $U(1)_Y $ & $\mathbb{Z}_2$\\
\hline
 $H_1$ & 1 & $\begin{pmatrix} \left(v_1+h_1+i \phi_1^0\right)/\sqrt{2} \\  \phi^- \end{pmatrix}$ & -1/2 & $+$ \\
 $H_2$ & 1 & $\begin{pmatrix} \phi^+ \\ \left(v_2+h_2+i \phi_2^0\right)/\sqrt{2} \end{pmatrix}$ & 1/2 & $-$ \\
 $Q$ &3 & $\begin{pmatrix} U \\ B \end{pmatrix}$ &  1/6 & $+$\\
$U'$ & 3& $U'$ & $2/3$ & $+$\\
 \hline
\end{tabular}
\end{center}
\label{4thgen}
\end{table}%
The type I 2HDM Lagrangian, in which the SM fermions couple only to $H_2$, is then augmented by the terms
\begin{align}\label{L}
	\mathcal{L} & \supset
	\left[ 
	y^L_Q \bar{Q}_L \tilde{H}_1 U'_R +y^R_Q \bar{Q}_R \tilde{H}_1 U'_L+
	\text{H.c.}\right]\nonumber\\
	& +
	m_Q \bar{Q} Q
	+		
	m_{U'} \bar{U}' U' \,,
\end{align}
plus additional mixing couplings with the SM quarks.  We do not write explicitly these terms which simply allow the vector-like quarks to decay to SM particles and avoid detection. We write in Appendix~\ref{msmx} the masses and relevant couplings of the mass eigenstates $T,\,T'$, and $B$. The experimental constraints from the processes $T,T'\rightarrow b W^+$ and $B\rightarrow b h$ require these masses to be larger than 705 and 846~GeV \cite{CMS:1900uua,CMS:2014afa}, respectively. To satisfy the experimental constraints we take the masses to be
\be
m_T=800~{\rm GeV}\,,\ m_{T'}=900~{\rm GeV}\,,\ m_{B}=850~{\rm GeV}\,,
\ee
and scan the full parameter space for data points producing a diphoton excess $\sigma_{pp\rightarrow H,A\rightarrow \gamma\gamma}=6$. To simplify the search we set $\tan\beta=6$, in a way that the SM fermion decay channels are highly suppressed. With our methodology we find a data point featuring a minimum average squared Yukawa coupling of
\be\label{viap}
m_{U}=755~{\rm GeV}\,,\, m_{Q}=850~{\rm GeV}\,,\, y_Q^L=10.3\,,\, y_Q^R=9.22\,.
\ee
Such point is phenomenologically viable, although the large Yukawa couplings  in Eqs.~\eqref{viap} are expected to drive the model to the non-perturbative regime at relatively low energy of $O({\rm TeV})$, close to the resonance mass \cite{Bertuzzo:2016fmv}.

\section{Generic technicolour} 
\label{sec:TC}

Finally, we would like to comment on the possibility that  the Higgs boson and the hinted new 750~GeV resonance be composite objects.
This scenario may be realised, for example, in a generic technicolour model. The 125~GeV Higgs in this case would be associated to a technidilaton, the composite pseudo-Nambu Goldstone boson of the approximate scale symmetry of the strongly coupled theory (see for example \cite{Yamawaki:1985zg,Appelquist:1998xf,Sannino:2004qp}, or \cite{Matsuzaki:2012xx,Belyaev:2013ida} for a study of the viability of these models at LHC, and \cite{Fukano:2015hga,Franzosi:2015zra} for an interpretation within the same frameworks of the diboson excess at LHC Run I). Other spin-zero resonances in this case would not be protected by such approximate symmetry, and their masses could be estimated by scaling up the corresponding QCD composite states via a straightforward dimensional analysis \cite{Manohar:1983md}. Assuming the 750~GeV resonance to be a CP-even state, a naive estimate of its mass is given by twice the mass of its techniquark constituents \cite{Hill:2002ap}
\be
m_H\sim 2\frac{m_P}{3} \frac{\sqrt{3}~v}{f_\pi \sqrt{N_D N_{\rm TC}}}=750~{\rm GeV}\ \Rightarrow \ N_D\sim \frac{12}{N_{\rm TC}},
\ee
with $P$ being the proton, $f_\pi=100$~MeV the QCD pion decay constant, $N_D$ the number of electroweak (EW) doublets, and $N_{\rm TC}$ the number of techni-colors. The equation above is satisfied for example by $N_{\rm TC}=4$ and three EW doublets. Another possibility is that $H$ is actually a composite pseudoscalar, corresponding to the QCD pion $\eta^\prime$: in this case a naive estimate based on the known QCD properties produces \cite{Hill:2002ap}
\be
m_H\sim m_{\eta^\prime}  \frac{3 \sqrt{18}~v}{2 f_\pi N_{\rm TC} \sqrt{N_D N_{\rm TC}}}=750~{\rm GeV}\ \Rightarrow \ N_D\sim \frac{400}{N^3_{\rm TC}}~.
\ee
In this case, again for $N_{\rm TC}=4$, the necessary number of EW doublets needed to explain the observed mass would be six. It is also worth to notice that these resonances, given their strong interactions to other composite states, are generally expected to have a wide width, which seems to be the case for the 750~GeV resonance observed at LHC.

In this scenario additional composite resonances, for example spin-one bosons $\rho_{\rm TC}$ with masses \cite{Hill:2002ap}
\be
m_{\rho TC}\sim m_\rho \frac{v \sqrt{3}}{\sqrt{N_D N_{\rm TC}}}~,
\ee 
of the order of several TeVs, could be within the reach of Run II LHC searches.
%

\section{Conclusions} 
\label{sec:conclusion}

We have argued in this work that the 750~GeV di-photon excesses seen by the ATLAS and the CMS collaborations may follow from the decays of a new resonance
with the global statistical significance exceeding $3~\sigma$.
We determined the cross sections of the signal times branching ratio in both experiments and find them to be consistent with each other at $1.8~\sigma$ level.
Using this result, we have shown that the di-photon excess can be explained consistently with the negative results for all other final states in the 
singlet scalar extensions of the SM and in 2HDM extended by two vector-like quarks. At the same time, the simplest 2HDMs and the MSSM, a UV completion of type II 2HDM, seem incompatible with the result. Consequently, 
in order to embed the observed phenomenology into a supersymmetric framework, non-standard extensions of the MSSM must be considered.
We finally commented on the possibility that the new hypothetical particle might be a spin zero resonance of some generic composite model and argued that in this scenario additional spin-one composite resonances would be within reach of Run II at LHC. While the LHC 750~GeV di-photon excess may still turn out to be a statistical fluctuation, we conclude that 
it is also a good and consistent candidate for the first signal of new physics beyond the SM.


\begin{acknowledgments}
The authors thank Mario Kadastik, Kristjan Kannike, Andrew Fowlie, Christian Spethmann, Christian Veelken and Hardi Veerm\"ae for useful discussions. 
This work was supported by the grants IUT23-6, PUTJD110, CERN+ and by EU through the ERDF CoE program.

\end{acknowledgments}
\onecolumngrid
\appendix
\section{Vector-like fermion masses and couplings}
\label{msmx}
The masses of the vector-like quark mass eigenstates $T,T',B$ are, respectively
\be\label{mx12}
m_{T,T'}=\frac{1}{2} \sqrt{l^2+L^2+m^2+M^2\mp2 \sqrt{\left(l^2+m^2\right) \left(L^2+M^2\right)}}\,,\,m_{B}=\frac{1}{2}\left(M+m\right)\,,
\ee
with
\be\label{MmLl}
M=m_Q+m_{U'}\,,\,m=m_{U'}-m_Q\,,\,L=-\frac{v_w \cos\beta}{\sqrt{2}}\left(y_Q^L+y_Q^R\right)\,,\, l=-\frac{v_w \cos\beta}{\sqrt{2}}\left(y_Q^L-y_Q^R\right) \,.
\ee
In terms of the same quantities, the coupling coefficients of $T,T'$ to the light Higgs $h$ are, respectively,
\be
a_{1,2}^h=\frac{L^2 m \left(m M\mp \sqrt{\left(l^2+m^2\right) \left(L^2+M^2\right)}\right)+l^2 \left[L^2 (m+M)+M \left(m M\mp \sqrt{\left(l^2+m^2\right)
   \left(L^2+M^2\right)}\right)\right]}{4 \left(l^2+m^2\right) \left(L^2+M^2\right) }  \frac{N_{1,2}}{m_{1,2}}\,,
\ee
while those to the heavy Higgs $H$ are
\be\label{yH}
a_{1,2}^H= \frac{ \left(l^2 M+L^2 m\right) \left(m M\mp \sqrt{\left(l^2+m^2\right) \left(L^2+M^2\right)}\right)+l^2 L^2 (m+M)}{4 \left(l^2+m^2\right)
   \left(L^2+M^2\right)}\frac{N_{1,2}}{m_{1,2}} \tan \beta\,,
   \ee
and those to the pseudoscalar $A$ are
\be\label{yA}
a_{1,2}^A= \frac{l L \left[(m+M) \left(m M\mp \sqrt{\left(l^2+m^2\right) \left(L^2+M^2\right)}\right)+l^2 M+L^2 m\right]}{4 \left(l^2+m^2\right) \left(L^2+M^2\right)}\frac{N_{1,2}}{m_{1,2}} \tan \beta\,,
   \ee
with $m_{1,2}=m_{T,T'}$ given by Eq.~\eqref{mx12}, and
\be
N_{1,2}=\sqrt{1+\left|\frac{l L-m M\mp \sqrt{\left(l^2+m^2\right) \left(L^2+M^2\right)}}{L m+l M}\right|^2} \sqrt{1+\left|\frac{l L+m M\pm
   \sqrt{\left(l^2+m^2\right) \left(L^2+M^2\right)}}{L m-l M}\right|^2}\,.
\ee
Finally, the relevant coupling coefficients of $B$ are all simply zero:
\be
a_B^h=a^H_B=a^A_B=0\,.
\ee

\twocolumngrid



\end{document}